\documentclass[11pt,a4paper]{article}
\usepackage{amsmath,amssymb,hyperref,epsfig,multicol}
\usepackage{graphicx}
\usepackage{amsmath}
\usepackage{amssymb}

\newcommand{\beq}{\begin{equation}}
\newcommand{\eeq}{\end{equation}}
\newcommand{\bea}{\begin{eqnarray}}
\newcommand{\eea}{\end{eqnarray}}

\newcommand{\be}{\begin{equation}}
\newcommand{\ee}{\end{equation}}

\newcommand{\cR}{{\mathcal R}}

\newcommand{\fL}{{\mathfrak L}}
\newcommand{\fR}{{\mathfrak R}}
\newcommand{\fQ}{{\mathfrak Q}}
\newcommand{\fS}{{\mathfrak S}}

\newcommand{\alg}[1]{\mathfrak{#1}}

\usepackage{epsfig,multicol}
\begin{document}

\begin{titlepage}
\begin{flushright} 
MIT-CTP 3953\\
\end{flushright}
\mbox{ }  \hfill 
\vspace{5ex}
\Large
\begin {center}     
{\bf Structure of the string R-matrix}
\end {center}
\large
\vspace{1ex}
\begin{center}
Alessandro Torrielli
\end{center}
\vspace{1ex}
\begin{center}
Center for Theoretical Physics\\
Laboratory for Nuclear Sciences\\
and\\
Department of Physics\\
Massachusetts Institute of Technology\\
Cambridge, Massachusetts 02139, USA\\ 
\vspace{1ex}
\texttt{torriell@mit.edu}

\end{center}
\vspace{4ex}
\rm
\begin{center}
{\bf Abstract}
\end{center} 
\normalsize 
By requiring invariance directly under the Yangian symmetry, we rederive
Beisert's quantum R-matrix, in a form that carries explicit dependence on the 
representation labels, the braiding factors, and the spectral parameters 
$u_i$. In this way, we demonstrate that there exist a rewriting of its entries,
such that
the dependence on the spectral parameters is purely {\it of difference form}.
Namely, the latter enter only in the combination $u_1-u_2$, as indicated by the shift automorphism of the Yangian. When 
recasted in this fashion, the entries exhibit a cleaner structure, which 
allows to 
spot new interesting relations among them. This permits to package them into a 
practical tensorial expression, where the non-diagonal entries are taken care 
by explicit combinations of symmetry algebra generators.

\vfill
\end{titlepage} 

\section{Introduction}
Integrability in AdS/CFT \cite{MZ,gauge,string,bl} (see \cite{rev} for reviews)
is strictly related to the existence of an
R-matrix \cite{B}, 
which is a solution of the quantum Yang-Baxter equation \cite{nlin,Gleb}, and 
satisfies the Hopf-algebraic analog of crossing symmetry \cite{J,GHPST}.  
This R-matrix exhibits a certain infinite-dimensional Yangian-type symmetry 
\cite{BYang},
based on the central extension of the Lie superalgebra $\alg{psu(2|2)}$,
plus an additional Yangian generator of $\alg{u(2|2)}$ signature 
\cite{MMT,BS}. A universal form of the R-matrix would be desirable, 
particularly in view of its potential use in studying finite-size effects 
\cite{finite1,finite2,finite3}.  
Such universal R-matrix is still unknown, mainly due to the 
fact that the dual Coxeter number of the algebra vanishes, and 
therefore traditional mathematical techniques do not 
straightforwardly apply \cite{Etingof,KT}. 
Nevertheless, huge progress has come in several 
aspects \cite{Yangian,Kazakov,Hubbard,qu,Moriyama}, like in the study of 
the correspondent classical r-matrix \cite{T,MT,BS,Moriyama,Leeuw}, 
in deriving higher representations \cite{GS}, and in giving the Yangian
an almost canonical form \cite{FA}.

The way the R-matrix was originally derived \cite{B} was by using the Lie
superalgebra symmetry in the fundamental representation. 
The combination of the peculiar features of the tensor products of two 
fundamentals \cite{nlin}, 
together with the non-trivial braiding of the coproduct 
\cite{GHPST,bl,Gleb}
allowed to fix all entries up to an 
overall scalar factor \cite{BHLBES,BCDKS}. Yangian symmetry was 
explicitly discovered 
only later \cite{BYang}, 
initially hardly visible due to the consistency 
relations that mix together all the 
parameters in the game. In higher
representations the Yangian 
may impose useful conditions \cite{Leeuw}, but the 
Yang-Baxter equation could probably be sufficient to fix the R-matrix 
\cite{GS}.
 
Were it not for the algebraic peculiarity mentioned above, 
then probably, as it happens for more traditional Yangians,
the Lie superalgebra symmetry 
would not be enough, and the invariance under the 
Yangian would determine much of the form of the R-matrix.
Since the Yangian carries the spectral parameter $u$, this normally 
gives the familiar
rational R-matrices, depending only on the difference $u_1 - u_2$ due to a constant
shift automorphism $u \rightarrow u+c$. 
From there, investigation of the universal R-matrix is easier. 

In this paper, we want to exploit this privileged viewpoint that the Yangian
gives on the R-matrix, and imagine we were to fix the latter by purely using 
the (first level) Yangian generators. Our strategy is to try to avoid explicit
use of the constraints, and to carry on the various parameters almost until
the end as if they were unrelated. 
In particular, the relevant Yangian still possesses the shift automorphism \cite{BYang,FA}.
Our calculation produces explicit expressions for the entries, where the dependence on the spectral 
parameters $u_i$ is indeed purely {\it of difference form}, namely $u_1 - u_2$, as expected for the above reasons. Moreover, the other parameters (representation labels, braiding factors) are also recognizable, which helps in finding new relations among
the entries, and tells us much about the combination of generators 
that is going to 
produce them. Upon imposing the constraints, we 
exactly recover the original quantum R-matrix found by Beisert.
Tempted by this new perspective, 
we also provide a tensorial repackaging of the 
entries which further clarifies their hidden structure, and might serve as
a device to investigate the properties of the universal R-matrix.

\section{Structural relations}
In this section, we will follow the approach of rederiving the quantum 
R-matrix, by 
using the 
fundamental representation not of the Lie superalgebra, as it is traditionally 
done, but rather using from the very beginning the first Yangian generators.
The correspondent 
coproducts are given in \cite{BYang,FA}. In doing this, we will
leave explicit the dependence on the representation labels, the braiding 
factors, and the spectral parameters, as if they were not related to each 
other. This will allow us to trace them back 
through the entries of the R-matrix.
Of course, only when they satisfy the familiar 
relations one can find a solution,
therefore it is impossible to really 
terminate the process without imposing those 
relations. Nevertheless, it is possible to proceed further enough to 
reach a satisfactory 
form for all the entries, which we report below. At that point, there is no 
need to continue, and we can just read the entries we obtain. 
As a check, we can then verify that, on the constraints, we recover
Beisert's quantum R-matrix.

Let us first remind the basic definition of the labels of the
fundamental representation: 

\begin{align}
&\fR^a{}_b|\phi^c\rangle=\delta^c_b|\phi^a\rangle - \frac{1}{2} \delta^a_b|\phi^c\rangle~,\quad 
\fL^\alpha{}_\beta|\phi^\gamma\rangle=\delta^\gamma_\beta|\phi^\alpha\rangle - \frac{1}{2} \delta^\alpha_\beta|\phi^\gamma\rangle~,\nonumber\\
&\fQ^\alpha{}_a|\phi^b\rangle= a \, \delta^b_a|\psi^\alpha\rangle~,\quad
\fQ^\alpha{}_a|\psi^\beta\rangle
= b \, \epsilon^{\alpha\beta}\epsilon_{ab}|\phi^b\rangle~,\nonumber\\
&\fS^a{}_\alpha|\phi^b\rangle
= c \, \epsilon^{ab}\epsilon_{\alpha\beta}|\psi^\beta\rangle~,\quad
\fS^a{}_\alpha|\psi^\beta\rangle= d \, \delta^\beta_\alpha|\phi^a\rangle~.
\end{align}
The labels satisfy $ad-bc=1$.
The central (braiding) operator $\alg{U}$ has eigenvalue $U$, and $u$ will be 
the spectral parameter.

Let us write the R-matrix in the familiar $\alg{su(2)}\oplus \alg{su(2)}$ 
invariant fashion: 

\begin{align}
\cR_{12}|\phi^a_1\phi^b_2\rangle
&=R^{12}_{12}|\phi^a_1\phi^b_2\rangle
+R^{12}_{21}|\phi^b_1\phi^a_2\rangle
+R^{12}_{34} \, \epsilon^{ab}\epsilon_{\alpha\beta}
|\psi^\alpha_1\psi^\beta_2\rangle~,\nonumber\\
\cR_{12}|\psi^\alpha_1\psi^\beta_2\rangle
&=R^{34}_{34}|\psi^\alpha_1\psi^\beta_2\rangle
+R^{34}_{43}|\psi^\beta_1\psi^\alpha_2\rangle
+R^{34}_{12} \, \epsilon^{\alpha\beta}\epsilon_{ab}
|\phi^a_1\phi^b_2\rangle~,\nonumber\\
\cR_{12}|\phi^a_1\psi^\beta_2\rangle
&=R^{13}_{13}|\phi^a_1\psi^\beta_2\rangle
+R^{13}_{31}\psi^\beta_1\phi^a_2\rangle~,\nonumber\\
\cR_{12}|\psi^\alpha_1\phi^b_2\rangle
&=R^{31}_{31}|\psi^\alpha_1\phi^b_2\rangle
+R^{31}_{13}|\phi^b_1\psi^\alpha_2\rangle~.
\end{align}
Choosing the overall normalization to be such that $R^{33}_{33} = R^{34}_{34} + R^{34}_{43}  \, = \, U_2/U_1$,  
the expressions for the entries which we obtain by imposing invariance under the 
Yangian are given 
by the following
formulas:

\begin{align}
\label{a}
&R^{31}_{31} = U_2 \, \frac{u_1 - u_2}{u_1 - u_2 - a_1 d_1 + 
a_1 b_1 (c_2/a_2) \, U_2^2}~,\nonumber\\
&R^{31}_{13} = R^{31}_{31} \, \frac{1}{U_2} \, \frac{(a_2 d_1 - 
b_1 c_2 \, U_2^2)}{u_1 - u_2}~,\nonumber\\
&R^{13}_{13} = \frac{U_2^2}{U_1} \, \frac{u_1 - u_2}{u_1 - u_2 + b_2 
c_2 - a_2 b_2 (d_1/b_1) \, U_1^2}~,\nonumber\\
&R^{13}_{31} = R^{13}_{13} \, \frac{1}{U_2} \, \frac{(- b_2 c_1 + 
a_1 d_2 \, U_2^2)}{u_1 - u_2}~,\nonumber\\
&R^{12}_{34}= R^{12}_{12} \, \frac{1}{U_2} \, \frac{(- c_1 a_2 + 
a_1 c_2 \, U_2^2)}{u_1 - u_2 - 1}~,\nonumber\\ 
&R^{12}_{21} = R^{12}_{12} \, \frac{1}{u_1 - u_2} \, \bigg( 1 + \frac{1}{U_2^2} \, \frac{(c_1 a_2 - 
a_1 c_2 \, U_2^2)(d_1 b_2 - 
b_1 d_2 \, U_2^2)}{u_1 - u_2 - 1}\bigg)~,\nonumber\\
&R^{34}_{12}= R^{34}_{34} \, \frac{1}{U_2} \, \frac{(d_1 b_2 - 
b_1 d_2 \, U_2^2)}{u_1 - u_2 + 1}~,\nonumber\\
&R^{34}_{43} = R^{34}_{34} \, \frac{1}{u_1 - u_2} \, \bigg(- 1 + \frac{1}{U_2^2}\, \frac{(c_1 a_2 - 
a_1 c_2 \, U_2^2)(d_1 b_2 - 
b_1 d_2 \, U_2^2)}{u_1 - u_2 + 1}\bigg)~, 
\end{align}

\begin{align}
\label{b}
&R^{11}_{11}= R^{12}_{12} +R^{12}_{21} = \frac{U_2^2}{U_1^2} \, \frac{u_1 - u_2 - b_1 c_1 + 
a_1 b_1 (d_2/b_2) \, U_2^2}{u_1 - u_2 + b_2 c_2 - 
a_2 b_2 (d_1/b_1) \, U_1^2}~,\\ 
&\nonumber
\end{align}

As one can immediately see, the dependence of the entries on the
spectral parameters $u_i$ is purely {\it of difference form}, as indicated by 
the fact that the Yangian is invariant under an automorphism that shifts the
spectral parameter by a constant. 

One can check that these expressions indeed reproduce the entries of Beisert's R-matrix \cite{B}. It is sufficient to recall the parametrization 
\begin{align}
\label{c}
a=\sqrt{g}\gamma~,\quad
b=\sqrt{g}\frac{\alpha}{\gamma}\biggl(1-\frac{x^+}{x^-}\biggr)~,\quad
c=\sqrt{g}\frac{i\gamma}{\alpha x^+}~,\quad
d=\sqrt{g}\frac{x^+}{i\gamma}\biggl(1-\frac{x^-}{x^+}\biggr)~,
\end{align}
together with $U=\sqrt{x^+/x^-}$ and the relation
$x^++1/x^+-x^--1/x^-=i/g$, and that $u = (ig/2)\, (x^+ + x^-)(1 + 1/x^+ x^-)$.
Then, we compare with the familiar formulas

\begin{align}
\cR_{12}|\phi^a_1\phi^b_2\rangle
&=\frac{1}{2}(A_{12}-B_{12})|\phi^a_1\phi^b_2\rangle
+\frac{1}{2}(A_{12}+B_{12})|\phi^b_1\phi^a_2\rangle
+\frac{1}{2}C_{12}\epsilon^{ab}\epsilon_{\alpha\beta}
|\psi^\alpha_1\psi^\beta_2\rangle~,\nonumber\\
\cR_{12}|\psi^\alpha_1\psi^\beta_2\rangle
&=-\frac{1}{2}(D_{12}-E_{12})|\psi^\alpha_1\psi^\beta_2\rangle
-\frac{1}{2}(D_{12}+E_{12})|\psi^\beta_1\psi^\alpha_2\rangle
-\frac{1}{2}F_{12}\epsilon^{\alpha\beta}\epsilon_{ab}
|\phi^a_1\phi^b_2\rangle~,\nonumber\\
\cR_{12}|\phi^a_1\psi^\beta_2\rangle
&=G_{12}|\phi^a_1\psi^\beta_2\rangle
+H_{12}|\psi^\beta_1\phi^a_2\rangle~,\nonumber\\
\cR_{12}|\psi^\alpha_1\phi^b_2\rangle
&=L_{12}|\psi^\alpha_1\phi^b_2\rangle
+K_{12}|\phi^b_1\psi^\alpha_2\rangle~,
\end{align}
with the functions $A_{12},B_{12},\ldots$ given by

\begin{align}
\label{d}
&A_{12}=\frac{x_2^+-x_1^-}{x_2^--x_1^+}~,\quad
B_{12}=\frac{x_2^+-x_1^-}{x_2^--x_1^+}
\biggl(1-2\frac{1-1/x_1^+x_2^-}{1-1/x_1^+x_2^+}
\frac{x_2^--x_1^-}{x_2^+-x_1^-}\biggr)~,\nonumber\\
&\qquad C_{12}=\frac{2\gamma_1\gamma_2U_2}{\alpha x_1^+x_2^+}
\frac{1}{1-1/x_1^+x_2^+}
\frac{x_2^--x_1^-}{x_2^--x_1^+}~,\nonumber\\
&D_{12}=-\frac{U_2}{U_1}~,\quad
E_{12}=-\frac{U_2}{U_1}
\biggl(1-2\frac{1-1/x_1^-x_2^+}{1-1/x_1^-x_2^-}
\frac{x_2^+-x_1^+}{x_2^--x_1^+}\biggr)~,\nonumber\\
&\qquad F_{12}=-\frac{2\alpha(x_1^+-x_1^-)(x_2^+-x_2^-)}
{\gamma_1\gamma_2U_1x_1^-x_2^-}
\frac{1}{1-1/x_1^-x_2^-}
\frac{x_2^+-x_1^+}{x_2^--x_1^+}~,\nonumber\\
&G_{12}=\frac{1}{U_1}\frac{x_2^+-x_1^+}{x_2^--x_1^+}~,\quad
H_{12}=\frac{\gamma_1U_2}{\gamma_2U_1}
\frac{x_2^+-x_2^-}{x_2^--x_1^+}~,\nonumber\\
&L_{12}=U_2\frac{x_2^--x_1^-}{x_2^--x_1^+}~,\quad
K_{12}=\frac{\gamma_2}{\gamma_1}
\frac{x_1^+-x_1^-}{x_2^--x_1^+}~.
\end{align}
By making use of (\ref{a}), (\ref{b}), (\ref{c}), (\ref{d}), it is possible to verify that the two 
R-matrices indeed exactly match.

As an interesting remark, we notice that in formulas (\ref{a}) and (\ref{b})
it is possible to {\it formally} switch off the braiding ($U_i\to 1$), 
and also
send everywhere  
$b_i$, $c_i$ to zero\footnote{Undetermined expressions like $b_1/b_2$ are 
sent to $1$.}, and $a_i$, $d_i$ to $1$. 
This should correspond to scatter two representations
of $\alg{gl(2|2)}$. Indeed, if one does that, one finds out that the R-matrix 
becomes formally equal to

\begin{equation}
\cR_{12}^{lim} = \frac{u_1 - u_2}{u_1 - u_2 - 1} \, \bigg( 1\otimes 1 
+ \frac{1}{u_1 - u_2} \sum_{i,j=1}^4 \, (-)^j \, E_{ij} \otimes E_{ji}\bigg)~,
\end{equation}
where $E_{ij}$ are the unit matrices with all zeroes but $1$ in position 
$(i,j)$, and bosonic and fermionic indices are altogether 
numbered from $1$ to $4$. The combination 
$\sum_{i,j=1}^4 \, (-)^j \, E_{ij} \otimes E_{ji}$ is the quadratic 
Casimir of $\alg{gl(2|2)}\otimes \alg{gl(2|2)}$, which already emerged from one-loop gauge theory \cite{gauge,finite1},
and from the classical
analysis of \cite{T}. The R-matrix which we obtain by this formal procedure is 
recognized this time as the 
quantum (Yang-type) R-matrix of $\alg{gl(2|2)}$ in the fundamental 
representation, and it is well known to solve the 
Yang-Baxter equation. We have thus find a consistent practical way of tuning 
on and off 
the central extensions
(which are proportional to $b$ and $c$ respectively) 
in the formula for the R-matrix.

\section{Tensorial repackaging}
We would like to exploit here the rewriting achieved in the previous section,
in order to express the whole R-matrix in a more compact tensorial form.
This will be far from enough to be able to
provide the universal R-matrix, but it will be
a useful exercise which may teach us where some of the terms are likely to
come from.
We (re)write here below 
some of the important identities that one obtains from the 
expressions given in the previous section:

\begin{align}
&R^{31}_{13} = R^{31}_{31} \, \frac{1}{U_2} \, \frac{(a_2 d_1 - 
b_1 c_2 \, U_2^2)}{u_1 - u_2}~,\nonumber\\
&R^{13}_{31} = R^{13}_{13} \, \frac{1}{U_2} \, \frac{(- b_2 c_1 + 
a_1 d_2 \, U_2^2)}{u_1 - u_2}~,\nonumber\\
&R^{12}_{34}= R^{12}_{12} \, \frac{1}{U_2} \, \frac{(- c_1 a_2 + 
a_1 c_2 \, U_2^2)}{u_1 - u_2} \, \frac{u_1 - u_2}{u_1 - u_2 - 1}~,\nonumber\\ 
&R^{34}_{12}= R^{34}_{34} \, \frac{1}{U_2} \, \frac{(d_1 b_2 - 
b_1 d_2 \, U_2^2)}{u_1 - u_2} \, \frac{u_1 - u_2}{u_1 - u_2 + 1}~,\nonumber\\
&R^{12}_{21} = R^{12}_{12} \, \frac{1}{U_2^2} \, \frac{(c_1 a_2 - 
a_1 c_2 \, U_2^2)}{u_1 - u_2} \, \frac{(d_1 b_2 - 
b_1 d_2 \, U_2^2)}{u_1 - u_2} \, \frac{u_1 - u_2}{u_1 - u_2 -1} + R^{12}_{12} \, \frac{1}{u_1 - u_2}~,\nonumber\\
&R^{34}_{43} = R^{34}_{34} \, \frac{1}{U_2^2} \, \frac{(c_1 a_2 - 
a_1 c_2 \, U_2^2)}{u_1 - u_2} \, \frac{(d_1 b_2 - 
b_1 d_2 \, U_2^2)}{u_1 - u_2} \, \frac{u_1 - u_2}{u_1 - u_2 +1} - R^{34}_{34} \, \frac{1}{u_1 - u_2}~.
\end{align}
This seems to suggest a not too unfamiliar 
matrix pattern. After trying different 
possibilities, one can see that a relatively simple 
tensorial repackaging of the R-matrix 
that is able to produce these relations can be found:

\begin{align}
&\cR_{12}= \cR^F_1 \, \cR^F_2 \, \cR^H_{12} + \cR^B_1 \, \cR^H_1 + \cR^B_2 \, \cR^H_2 ~,    
\end{align}
where 
\begin{align}
&\cR^F_1 = 1 \otimes 1 + \frac{1}{u_1 - u_2} \, (\fQ^1{}_1 \otimes \fS^1{}_1 \alg{U} + 
\fQ^2{}_2 \otimes \fS^2{}_2 \alg{U} - \fS^1{}_1 \otimes \fQ^1{}_1 \alg{U}^{-1} - 
\fS^2{}_2 \otimes \fQ^2{}_2 \alg{U}^{-1})~,\nonumber\\
&\cR^F_2 = 1 \otimes 1 + \frac{1}{u_1 - u_2} \, (\fQ^1{}_2 \otimes \fS^2{}_1 \alg{U} + 
\fQ^2{}_1 \otimes \fS^1{}_2 \alg{U} - \fS^1{}_2 \otimes \fQ^2{}_1 \alg{U}^{-1} - 
\fS^2{}_1 \otimes \fQ^1{}_2 \alg{U}^{-1})~,\nonumber\\
&\nonumber\\
&\cR^B_1 = \frac{\Pi_B \otimes \Pi_B}{1 - u_1 + u_2} + \frac{1}{u_1 - u_2} \, (\fR^1{}_2 \otimes \fR^2{}_1 + 
\fR^2{}_1 \otimes \fR^1{}_2)~, \quad\quad \quad \cR^H_1 = R^{12}_{12} \, \, 1\otimes 1~,\nonumber\\
&\cR^B_2 = \frac{\Pi_F \otimes \Pi_F}{1 + u_1 - u_2} - \frac{1}{u_1 - u_2} \, (\fL^1{}_2 \otimes \fL^2{}_1 + 
\fL^2{}_1 \otimes \fL^1{}_2)~, \quad\quad \quad \, \, \, \, \cR^H_2 = R^{34}_{34} \, \, 1\otimes 1~,\nonumber\\
&\nonumber\\
&\cR^H_{12} = R^{13}_{13} \, \, \Pi_B \otimes \Pi_F \, + R^{31}_{31} \, \, \Pi_F \otimes \Pi_B \, +  \cR^H_B~,\\ \nonumber
\end{align}
$\Pi_B$ and $\Pi_F$ being the projectors onto the 
bosonic and fermionic subspaces respectively, $\Pi_B = diag\{1,1,0,0\}$ and 
$\Pi_F = diag\{0,0,1,1\}$. 
The only non-zero entries of $\cR^H_B$ are
\begin{align}
&\cR^H_B |\phi^a_1 \phi^a_2 \rangle
=\bigg( R^{11}_{11} + \frac{R^{12}_{12}}{u_1 - u_2 - 1} \bigg)\, |\phi^a_1 \phi^a_2 \rangle~,\quad \, \, \, \, 
\cR^H_B \, \epsilon_{a b} |\phi^a_1 \phi^b_2 \rangle= R^{12}_{12} \, \frac{u_1 - u_2}{u_1 - u_2 - 1} \, \epsilon_{a b} |\phi^a_1 \phi^b_2 \rangle~,\nonumber\\
&R^H_B |\psi^\alpha_1 \psi^\alpha_2 \rangle
=\bigg( R^{33}_{33} - \frac{R^{34}_{34}}{u_1 - u_2 + 1} \bigg)\, |\psi^\alpha_1 \psi^\alpha_2 \rangle~,\quad
\cR^H_B \, \epsilon_{\alpha \beta} |\psi^\alpha_1 \psi^\beta_2 \rangle= R^{34}_{34} \, \frac{u_1 - u_2}{u_1 - u_2 + 1} \,\epsilon_{\alpha \beta } |\psi^\alpha_1 \psi^\beta_2 \rangle~.
\end{align}

This has to be seen as a practical tool to organize the entries, useful for a 
subsequent attempt to find the universal R-matrix. The Cartan part is 
traditionally the most complicated to reproduce, as shown by 
formulas (\ref{a}), (\ref{b}).
But also the root part seems to be a little different from standard cases, 
even when taking into account that in the fundamental representation the roots 
are 
nilpotent, and therefore what appears to be 
linear could well be the first term of a series expansion. 
Nevertheless, in the classical limit,
the formula seems to reproduce the non-diagonal part of the 
classical r-matrix. 
In any case, no conclusions are to be drawn at the moment on the 
persistence of 
this kind of pattern at the universal level, and the 
device is purely practical.

\section{Conclusions}
In this paper we have shown that there exist a way of rewriting Beisert's 
quantum 
R-matrix, such that the dependence on the spectral parameters is purely of 
difference form. This was hidden before by the complicated relations 
connecting the spectral parameter with the representation labels and the braiding factors, and can be
achieved by using the Yangian symmetry. In particular, this dependence
is expected from the shift 
automorphism that the Yangian possesses. The remaining dependence on the 
two representations, which is responsible for the ultimately observed 
non-difference pattern, is encoded in a precise sequence of
representation labels and braiding 
factors. This, in turn, is to be expected, if the R-matrix has to come from some 
universal combination of symmetry generators. We provide some hints at this 
structure, suggested by novel relations among the entries, which in this 
new rewriting are easier to spot.

When comparing with the classical r-matrix analysis of \cite{BS}, one notices
that a suitable readjusting of the classical
variables can be used to abandon a pure $u_1 - u_2$ dependence on the classical
spectral parameters, in order to reach 
a convenient normalization for the extra generator to insert in the classical
Yangian.
The quantum
version of this generator \cite{MMT,BS}
has not yet been canonically embedded in the quantum 
Yangian,
and it is hard to judge on its role at the moment. In other words, it is 
possible that one may 
have to eventually introduced a few uncoupled $u_i$'s in order to 
get a truly universal expression. What we have shown here is that, 
nevertheless, {\it there exist} 
at least one rewriting where the spectral parameters come in 
differences, all the rest being taken care of by recognizable algebraic 
quantities.

\section{Acknowledgments}
The author would like to thank Pavel Etingof and Victor Kac 
for discussion.
Many of the calculations have been performed with \texttt{Mathematica}.
This work is supported in part by funds provided by the U.S. 
Department of Energy (D.O.E.) under cooperative research agreement
DE-FG02-05ER41360. The author thanks Istituto Nazionale di Fisica Nucleare
(I.N.F.N.) for supporting him through a ``Bruno Rossi'' postdoctoral 
fellowship.

\end{document}